\documentclass[12pt]{article}
\usepackage{amsfonts,amsmath,amssymb}
\usepackage{graphics,epsfig}
\usepackage{color}

\numberwithin{equation}{section}
\definecolor{darkblue}{rgb}{0.02,0.608,0.808}
\definecolor{darkgreen}{rgb}{0,0.35,0}

\begin{document}

\title{Topological black holes in pure Gauss-Bonnet gravity and phase transitions}
\author{Ligeia Ar\'anguiz$^{1,2}$\thanks{ligeia.aranguiz@postgrado.usm.cl}\;,
Xiao-Mei Kuang$^{1,3}$\thanks{xiaomei.kuang@ucv.cl}\;
and Olivera Miskovic$^1$\thanks{olivera.miskovic@pucv.cl} \bigskip \\
{\small {$^1$Instituto de F\'isica, Pontificia Universidad Cat\'olica de
Valpara\'iso,}}\\
{\small {Casilla 4059, Valpara\'iso, Chile}}\\
{\small {\ $^2$Universidad T\'ecnica Federico Santa Mar\'\i a,} {Casilla
110-V, Valpara\'{\i}so, Chile}}\\
{\small {$^3$Department of Physics, National Technical University of Athens,}}\\
{\small {GR-15780 Athens, Greece}}}
\maketitle

\begin{abstract}
We study charged, static, topological black holes in pure Gauss-Bonnet gravity in asymptotically AdS space. As in general relativity, the theory possesses a unique nondegenerate AdS vacuum. It also admits charged black hole solutions which asymptotically behave as the Reissner-Nordstr\"{o}m AdS black hole.  We discuss black hole thermodynamics of these black holes. Then we study phase transitions in a dual quantum field theory in four dimensions, with the St\"{u}ckelberg scalar field as an order parameter. We find in the probe limit that the black hole can develop hair below some critical temperature, which suggests a phase transition. Depending on the scalar coupling constants, the phase transition can be first or second order. Analysis of the free energy reveals that, comparing the two solutions,  the hairy state is energetically favorable, thus a phase transition will occur in a dual field theory.
\end{abstract}

\section{Introduction}

It has been generally accepted that a higher-derivative gravity provides
corrections to general relativity, especially in the framework of AdS/CFT
correspondence \cite{Maldacena:1997re,Gubser:2002,Witten:1998}, where large $N$
corrections are of this form in a\ dual quantum field theory. If one
requires that the equations of motion contain at most second order
derivatives, its most general form in higher dimensions is given by the
Lovelock-Lanczos action \cite{Lovelock1,Lovelock2}. It is a sum of
dimensionally continued Euler densities, each one of order $p\geq 0$ in the
curvature, and is characterized by a set of the coupling constants
$\{\alpha_{p}\}$. This generalization of gravity has some distinguished properties.
The higher-order terms contribute to the dynamics only in $D\geq 2p+1$
dimensions and they are free of ghosts in the flat background \cite{Boulware:1985wk}.
They also possess black hole solutions whose
 thermodynamic properties have been extensively studied in the literature,
for example, in the Einstein-Gauss-Bonnet case \cite{Cai:2001dz} and, more
generally, Chern-Simons gravity \cite{Banados:1993ur} and different Lovelock
theories \cite{Myers:1988ze,Cai:2003kt,Garraffo:2008hu}. The coupling
constants can be fixed requiring a unique vacuum in the theory
\cite{Crisostomo:2000bb}. One way to have a sensible theory is to ensure a single
vacuum in any dimension. In that way, one obtains the Einstein-Hilbert
action, with only the linear term in the curvature nonvanishing (the
coupling $\alpha _{1}$) and the cosmological constant ($\alpha _{0}$). When
higher-order terms are present, it gives rise to Chern-Simons gravity
\cite{Chamseddine:1989nu} in odd dimensions and Born-Infeld gravity
\cite{Banados:1993ur} in even dimensions; but these vacua are $[(D-1)/2]$-fold
degenerate, which does not allow a perturbative analysis around it.

Another way to have higher-order curvature terms and the unique vacuum in Lovelock gravity
which is not degenerate, is to keep only one Euler term of order $p$ in
Riemann curvature and the cosmological constant; then one obtains pure
Lovelock gravity \cite{Cai:2006pq,Dadhich:2012ma}. This gravity possesses
black hole solutions which asymptotically coincide with the corresponding
Einstein solution in spite of the more complicated field equations
\cite{Cai:2006pq,Dadhich:2012ma}. It has also been  noted that a particular class
of pure Lovelock black holes with the maximal order of curvature, $p=[(D-1)/2]$,
has thermodynamical parameters with the universal behavior in terms of the
event horizon radius \cite{Dadhich:2012eg}. Interestingly, black
strings and branes cannot be constructed in general Lovelock theories, but
they can in a pure Lovelock case \cite{Kastor:2006vw,Giribet:2006ec}.

This intriguing behavior of pure Lovelock black holes motivates us to use
the AdS/CFT prescription to study a holographic quantum field theory dual to
pure Lovelock gravity. AdS/CFT duality links an asymptotically anti-de Sitter(AdS)
gravitational bulk theory to a dual quantum theory living on its boundary.
Recent progress on this subject indicates that the duality plays an
important role in studying various strongly coupled phenomena in condensed matter physics
\cite{Hartnoll:2009sz}, especially in building a gravitational dual model
for a superconductor with either an electric \cite{Gregory:2009fj} or
a magnetic field \cite{Albash:2009ix,Ge:2010aa,Domenech:2010nf,Kuang:2013oqa}
(see review papers \cite{Hartnoll:2008kx,Herzog:2009xv}).

Our motivation allows us to benefit in two aspects. On one hand, the
presence of higher-order terms in gravity implies, in the context of AdS/CFT
correspondence, an appearance of new couplings among quantum operators in a
holographic conformal field theory. Thus, higher-curvature interaction in
pure Lovelock gravity is also expected to show new features in a dual field
theory. It has already been observed that there are holographic s-wave
\cite{Hartnoll:2008vx} and p-wave \cite{Gubser:2008wv,Cai:2010cv} phase
transitions in a superconductor dual to higher-order gravity, such as
Einstein-Gauss-Bonnet superconductor \cite{Pan:2009xa,Pan:2010at,Barclay:2010up},
and the field theories dual to quasitopological gravity
\cite{Kuang:2010jc,Kuang:2011dy}, as well as
Lovelock gravity \cite{Aranguiz:2013cwa,Lin:2014vla}. Numerous work on this
topic confirms that higher-order terms indeed have a notable effect on
phase transitions, as they modify previously universal behavior of
holographic theories. On the other hand, the study of phase transitions of
a superconductor via AdS/CFT duality can show some insight into the (in)stabilities
of black holes. In particular, the stability of black holes in pure Lovelock
gravity has been discussed in \cite{Gannouji:2013eka}.

In this paper, we first add an electric charge to the pure Lovelock solutions
\cite{Cai:2006pq,Dadhich:2012ma} and we obtain charged black
holes whose asymptotic behavior is similar to the one of the
Reissner-Nordstr\"{o}m (A)dS black holes. In the AdS case, only topological black holes with hyperbolic
horizons will form. Then we focus on the simplest case of pure Gauss-Bonnet
AdS gravity in five dimensions and couple it to the electromagnetic
and massive St\"{u}ckelberg scalar fields. We explore the thermodynamics of the
black holes and the possible phase transitions affected by the St\"{u}ckelberg
correction with backreaction. The physical explanation of condensation in this kind of superconductors
is the breaking of the Abelian-Higgs mechanism. A gapless superconductor with
hyperbolic geometry has been studied in Ref.~\cite{Koutsoumbas:2009pa},
in which the mechanism of condensation is due to the coupling.

The rest of the paper is organized as follows. In Section 2, we review the
pure Lovelock gravity and obtain a charged black hole solution. Then we
introduce a holographic setup of a superconductor by coupling charged pure
Gauss-Bonnet AdS gravity with a St\"{u}ckelberg complex scalar in Section 3.
Next, the black hole thermodynamics including the Gibbs free energy,
conserved charges and quantum statistical relation are analytically
investigated. Section 5 shows our numerical results of
two phases and the free energy in the case of a probe limit.
We close with the conclusions and discussion.

\section{pure Lovelock gravity}

pure Lovelock (PL) gravity action in $d+1$ dimensions possesses, apart from
the cosmological constant, a kinetic term which is a single Lovelock
term of order $p$ in the Riemann curvature \cite{Dadhich:2012ma},
\begin{equation}
I_{\text{PL}}=-\frac{1}{2\kappa}\int d^{d+1}x\sqrt{-g}\left( \frac{1%
}{2^{p}}\ \delta _{\nu _{1}\cdots \nu _{2p}}^{\mu _{1}\cdots \mu
_{2p}}\,R_{\mu _{1}\mu _{2}}^{\nu _{1}\nu _{2}}\cdots R_{\mu _{2p-1}\mu
_{2p}}^{\nu _{2p-1}\nu _{2p}}-2\Lambda \right) \,.  \label{Eq-actionPL}
\end{equation}%
A polynomial in the curvature is a $2p$-dimensional Euler
invariant continued to $d+1$ dimensions. The tensor $\delta _{\nu
_{1}\cdots \nu _{2p}}^{\mu _{1}\cdots \mu _{2p}}=\det \left[ \delta _{\nu
_{1}}^{\mu _{1}}\cdots \delta _{\nu _{2p}}^{\mu _{2p}}\right] $ is the
antisymmetric Kronecker delta of rank $2p$ and $\Lambda $ is the
cosmological constant whose units are (length)$^{-2p}$, and the gravitational
constant $\kappa$ has units (length)$^{d+1-2p}$. In our notation, the
metric field $g_{\mu \nu }(x)$ is mostly positive and the Riemann curvature
reads $R_{\;\;\nu \alpha \beta }^{\mu }=\partial _{\alpha }\Gamma _{\nu
\beta }^{\mu }-\partial _{\beta }\Gamma _{\nu \alpha }^{\mu }+\Gamma
_{\lambda \alpha }^{\mu }\Gamma _{\nu \beta }^{\lambda }-\Gamma _{\lambda
\beta }^{\mu }\Gamma _{\nu \alpha }^{\lambda }$.

When $p=1$, the action describes Einstein-Hilbert gravity with the (usual)
cosmological constant of dimension (length)$^{-2}$. When $2\leq p\leq \left[
d/2\right] $, the theory becomes pure Lovelock gravity. Thus, the simplest theory of
this type is pure Gauss-Bonnet gravity in five dimensions which contains the
Gauss-Bonnet term $R^{2}-4R_{\mu \nu }R^{\mu \nu }+R_{\mu \nu \lambda \sigma
}R^{\mu \nu \lambda \sigma }$.

Equations of motion in PL gravity read
\begin{equation}
^{(p)}G_{\nu }^{\mu }\equiv -\frac{1}{2^{p+1}}\ \delta _{\nu \nu _{1}\cdots
\nu _{2p}}^{\mu \mu _{1}\cdots \mu _{2p}}\,R_{\mu _{1}\mu _{2}}^{\nu _{1}\nu
_{2}}\cdots R_{\mu _{2p-1}\mu _{2p}}^{\nu _{2p-1}\nu _{2p}}+\delta _{\nu
}^{\mu }\,\Lambda =0\,,  \label{(p)G}
\end{equation}
where $^{(p)}G_{\mu \nu }$ is a generalized Einstein tensor. As in any
Lovelock gravity, they are second order field equations in the metric.

A particular solution of these equations is the maximally symmetric
spacetime with constant scalar curvature. This is flat space when $\Lambda
=0 $ and $R_{\alpha \beta }^{\mu \nu }=\pm \frac{1}{\ell ^{2}}\,\delta
_{\alpha \beta }^{\mu \nu }$ when $\Lambda \neq 0$, corresponding to dS, (sign $+$) or AdS space (sign $-$). The
effective (A)dS radius $\ell $ is related to a nonvanishing cosmological
constant as
\begin{equation}
\Lambda =\frac{\left( \pm 1\right) ^{p}\,d!}{2(d-2p)!\ell ^{2p}}\,.  \label{Eq-l}
\end{equation}
Note that (A)dS space is not directly related to the sign of the cosmological
constant, as happens in general relativity, because the definition of
(A)dS space is associated with the sign of curvature, and not of $\Lambda$.
In five dimensions, for example, pure Gauss-Bonnet gravity ($p=2$)
has a positive cosmological constant, $\Lambda =12/\ell ^{4}$, and the
curvature of the maximally symmetric vacuum can be either positive or
negative. Indeed, writing the generalized Einstein tensor (\ref{(p)G}) in
the form%
\begin{equation}\label{eq-G2}
^{(2)}G_{\nu }^{\mu }=-\frac{1}{8}\ \delta _{\nu \nu _{1}\cdots \nu
_{4}}^{\mu \mu _{1}\cdots \mu _{4}}\,\left( R_{\mu _{1}\mu _{2}}^{\nu
_{1}\nu _{2}}-\frac{1}{\ell ^{2}}\,\delta _{\mu _{1}\mu _{2}}^{\nu _{1}\nu
_{2}}\right) \left( R_{\mu _{3}\mu _{4}}^{\nu _{3}\nu _{4}}+\frac{1}{\ell
^{2}}\,\delta _{\mu _{1}\mu _{2}}^{\nu _{1}\nu _{2}}\right),
\end{equation}%
it is clear that the vacuum $^{(2)}G_{\nu }^{\mu }=0$ can be in either dS
or AdS space. In general, for even $p$, the generalized Einstein tensor always has the form
$^{(p)}G=\left( R-\frac{1}{\ell ^{2}}\,\delta ^{2}\right) \left( R+
\frac{1}{\ell ^{2}}\,\delta ^{2}\right) \mathcal{P}(R)$, where the
polynomial $\mathcal{P}$ in the curvature does not have real
roots. When $p$ is odd, then $^{(p)}G=\left( R\pm \frac{1}{\ell ^{2}}\,\delta ^{2}\right) \mathcal{P}(R)$
has exactly one real root. Thus, there is always at most one (A)dS vacuum, and it is not
degenerate, which is suitable to study a class of asymptotically (A)dS spacetimes.

We show next that these spacetimes possess static charged black hole
solutions. We focus on the case with $\Lambda \neq 0$.

\subsection{Exact charged black hole solutions}

Consider the pure Lovelock action coupled to a Maxwell field $A_{\mu }(x)$ whose field strength is $F_{\mu \nu }=\partial _{\mu }A_{\nu }-\partial _{\nu
}A_{\mu }$,
\begin{equation}
I_{\text{PL}}+I_{\text{M}}=-\frac{1}{2\kappa}\int d^{d+1}x\sqrt{-g}%
\left( \frac{1}{2^{p}}\ \delta _{\nu _{1}\cdots \nu _{2p}}^{\mu _{1}\cdots
\mu _{2p}}\,R_{\mu _{1}\mu _{2}}^{\nu _{1}\nu _{2}}\cdots R_{\mu _{2p-1}\mu
_{2p}}^{\nu _{2p-1}\nu _{2p}}-2\Lambda -\frac{1}{4e^{2}}\,F^{2}\right) .
\end{equation}%
A matter source in the gravitational equations of motion is the electromagnetic
energy-momentum tensor,
\begin{equation}
T_{\mu \nu }=\frac{1}{2e^{2}}\,\left( F_{\mu \lambda }F_{\nu }^{\ \lambda }-%
\frac{1}{4}\,g_{\mu \nu }\,F^{2}\right) \,,
\end{equation}%
and the field equations read
\begin{eqnarray}
^{(p)}G_{\mu \nu } &=&T_{\mu \nu }\,,  \notag \\
\nabla _{\mu }F^{\mu \nu } &=&0\,.
\end{eqnarray}%
We take a static and spherically symmetric ansatz for the Maxwell field, $%
A_{\mu }=\delta _{\mu }^{t}\,\phi (r)$, as well as for the metric $g_{\mu
\nu }$,
\begin{equation}
ds^{2}=-f(r)\,dt^{2}+\frac{dr^{2}}{f(r)}+r^{2}d\Omega _{d-1}^{2}\,,
\label{static no hair}
\end{equation}%
where the transversal section $d\Omega _{d-1}^{2}$ is the maximally symmetric space of the unit
radius whose curvature is $k=1$ for dS space or $k=0,\pm 1$ for AdS space. 

It is worth noticing that the most general spherically symmetric ansatz possesses two independent metric functions 
$g_{tt} = -f(r)$ and $g_{rr} = 1/N (r)f (r)$ instead of (\ref{static no hair}) and that, in higher-order gravities, it cannot be taken for granted that field equations would uniquely solve them. In Lovelock gravity this happens only when the couplings are such that the theory possesses a degenerate vacuum \cite{Wheeler}, as, for example, in Chern-Simons (super-)gravity \cite{Zanelli}. In our case, however, it is clear from Eq.(\ref{eq-G2}) and the discussion below it, that the vacuum is always nondegenerate, so $g_{tt}(r)$ and $g_{rr}(r)$ are dynamical functions. This will be explicitly shown in Sec. 3.1.

In the ansatz (\ref{static no hair}), the equations of motion become
\begin{eqnarray}
0 &=&\frac{(-1)^{p-1}(d-1)!}{2(d-2p)!r^{2p}}\,(f-k)^{p-1}\left[ prf^{\prime
}+(d-2p)(f-k)\right] +\Lambda +\frac{\phi ^{\prime 2}}{4e^{2}}\,,  \notag \\
0 &=&\phi ^{\prime \prime }+\frac{d-1}{r}\,\phi ^{\prime }\,,
\end{eqnarray}%
where the prime stands for $d/dr$. A solution for the electric potential is
\begin{equation}
\phi (r)=\mu -\frac{\rho }{r^{d-2}}\,,
\end{equation}%
where $\mu =\phi (\infty )$ is the chemical potential and $\rho $ is an
integration constant related to the electric charge. The metric function
satisfies
\begin{equation}
(f-k)^{p}=\left( \mp \frac{r^{2}}{\ell ^{2}}\right) ^{p}-\frac{M_{0}}{%
r^{d-2p}}-\frac{\rho ^{2}}{c\,e^{2}r^{2d-2p-2}}\,,  \label{Eq-f-k}
\end{equation}%
where $M_{0}$ is an integration constant related to the mass and $c=\frac{%
2(-1)^{p}(d-1)(d-3)!}{(d-2p)!}$ is a number.

Taking the $p-$th root of the above equation leads to
\begin{equation}
f(r)=k\mp \frac{r^2}{\ell ^2}\left[ 1-(\mp 1)^p\,\left(\frac{M_0\ell ^{2p}}{%
r^d} +\frac{\rho^2\ell^{2p}}{ce^2r^{2d-2}}\right) \right] ^{\frac{1}{p}}\,.
\label{Eq-f}
\end{equation}
The scalar curvature behaves as $R\sim (M_0/r^d)^{1/p}$, so the spacetime of the form (\ref{Eq-f}) has a singularity in the origin, $r=0$, when $M_0
\neq 0$ and $p>0$.  To avoid a naked singularity, we require that it is protected by
the black hole horizon, $r_{+}$, which is the largest root of the equation $f(r_+)=0$,
and also that $f(r)\geq 0$ outside the black hole ($r\geq r_{+}$).

In dS space, the transversal section is always spherical, $k=1$, and the black hole generally exists. In AdS space,
only black holes with flat ($k=0$) and hyperbolic ($k=-1$) horizons are
formed.

In the asymptotic region ($r\rightarrow \infty $), without electric charge ($\rho =0$),
a black hole behaves in the leading order as the
Schwarzschild-(A)dS with the mass $M_{S}=(\mp 1)^{p-1}M_{0}\ell ^{2p-2}/p$.
Turning on the electric charge, a black hole with $\Lambda \neq 0$ is an
asymptotically Reissner-Nordstr\"{o}m (RN) (A)dS with the charge $Q_{RN}^{2}=\rho ^{2}\ell ^{2p-2}/e^{2}p$,
\begin{equation}
f\sim k\mp \frac{r^{2}}{\ell ^{2}}-\frac{M_{S}}{r^{d-2}}+\frac{Q_{RN}^{2}}{%
cr^{2d-4}}\,.  \label{f_asympt}
\end{equation}
This solution is a generalization of the black holes discussed in Refs.~\cite{Cai:2006pq,Dadhich:2012ma}
to the electrically charged ones with nonspherical horizons. Since the sign of mass $M_{S}$ can be positive or negative, and also the electrostatic
energy can decrease the total energy of the black
hole (when $c$ is negative), it may not be stable. Thermodynamic stability of neutral pure Lovelock
black holes has been discussed in Ref.~\cite{Cai:2006pq} and recently in Ref.~\cite{Gannouji:2013eka}.

In what follows, we explore thermal field theories dual to PL black holes in the
framework of gauge/gravity duality. We point out that, even though asymptotic
behaviors of Schwarzschild and PL black holes are similar, the dynamics of
respective spacetimes are different and their holographic theories will be
different, as well.

\subsubsection{Black holes in asymptotically AdS space \label{AAdS}}

For studying thermal field theories dual to PL gravity, we
are interested in AdS black holes (\ref{Eq-f}) with noncompact horizons $k=0$ and $k=-1$,
\begin{equation}
f(r)=k+\frac{r^{2}}{\ell ^{2}}\left( 1-\frac{M_{0}\ell ^{2p}}{r^{d}}-\frac{%
\rho ^{2}\ell ^{2p}}{ce^{2}r^{2d-2}}\right) ^{\frac{1}{p}}.  \label{f_d,p}
\end{equation}

Let us first consider the planar case with $k=0$. Without electric charge, the
horizon $r_+=(M_0\ell^{2p}) ^{\frac{1}{d}}$ is formed when $M_0>0$. However,
in that case the Hawking temperature becomes infinite for PL black holes ($p>1$),
\begin{equation}
T=\frac{f^{\prime }(r_+)}{4\pi }\sim \frac{1}{f(r_+)^{p-1}}
\rightarrow \infty \,. \label{planarT}
\end{equation}
Infinite temperature is due to the fact that the scalar curvature is singular on the horizon,
because we have $R\propto f(r_{+})^{1-2p}$. In that case the temperature formula (\ref{planarT}) might not  be applicable, as it considers only the singularity at $r=0$. Since the spacetimes with singularity horizons are not described within  the standard
framework of AdS/CFT correspondence in asymptotically AdS spaces, we are not
interested in these cases.

Another  possibility for having a noncompact horizon is to look at a hyperbolic
geometry whose horizon curvature is $k=-1$. Then a neutral black hole has a
horizon that satisfies the equation $f(r_{+})=0$, or
\begin{equation}
r_{+}^{d}-r_{+}^{d-2p}-M_{0}\ell ^{2p}=0\,.
\end{equation}%
An existence of $r_{+}$ depends on dimension, $M_{0}$ and $p$. When it
exists, the temperature is finite and it behaves as
\begin{equation}
T\sim r_{+}^{2p-1}\,,
\end{equation}%
which is suitable for holographic studies and high temperatures correspond
to large black holes.

Adding the electric charge to the black hole is equivalent to the shift $%
M_{0}\rightarrow M_{0}+\rho ^{2}/ce^{2}r^{d-2}$. The horizon again exists
when $k=-1$ and it satisfies the equation
\begin{equation}
\frac{r_{+}^{2d-2}}{\ell ^{2p}}-r_{+}^{2d-2p-2}-M_{0}r_{+}^{d-2}-\frac{\rho
^{2}}{ce^{2}}=0\,.  \label{horizon(d,p)}
\end{equation}%
Since the only condition is $d\geq 2p$, solutions of the above polynomial
depend on particular values of the coefficients and degree of the polynomial
and should be solved case by case.

On the other hand, electric properties of the charged black holes are described by the electric
potential. Regularity of the potential on the horizon requires $\phi
(r_{+})=0$, which relates the electric charge with the chemical potential as
$\rho =\mu r_{+}^{d-2}$, leading to%
\begin{equation}
\phi (r)=\frac{\rho }{r_{+}^{d-2}}\,\left( 1-\frac{r_{+}^{d-2}}{r^{d-2}}%
\right) \,.
\end{equation}

\subsubsection{pure Gauss-Bonnet AdS black hole \label{PGB AdS black hole}}

\begin{figure}[h]
\center{
\includegraphics[scale=0.8]{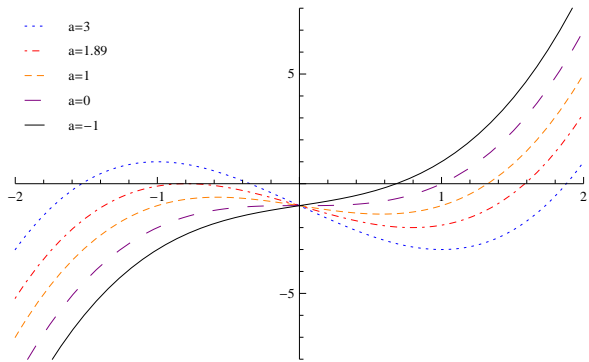}
\caption{\label{fig-horizon} Behaviour of the third order polynomial $x^{3}-ax-1=0$ in terms of the variable
$x=\left( \frac{\rho ^{2}\ell ^{4}}{6e^{2}}\right) ^{-1/3}r_{+}^{2}$
for different values of the parameter $a=\left( M_{0}+1\right)
\ell ^{4}\left( \frac{\rho ^{2}\ell ^{4}}{6e^{2}}\right) ^{-2/3}$. There is always only one positive
zero $x_+$ that corresponds to a unique (real) black hole horizon $r_+$.
}}
\end{figure}

Consider the simplest case of pure Gauss-Bonnet (PGB) gravity in five
dimensions, with the Gauss-Bonnet term as the kinetic term and the
cosmological constant $\Lambda =12/\ell ^{4}$. The PGB action with $p=2$ in
Eq. (\ref{Eq-actionPL}) is
\begin{equation}
I_{\text{G}}=-\frac{1}{2\kappa }\int d^{5}x\sqrt{-g}\,\left( R^{\mu \nu
\alpha \beta }R_{\mu \nu \alpha \beta }-4R^{\mu \nu }R_{\mu \nu
}+R^{2}-2\Lambda \right) \,,  \label{PGB}
\end{equation}%
and we couple it to the Maxwell field. An AdS solution for the metric
function is
\begin{equation}
f(r)=-1+\frac{r^{2}}{\ell ^{2}}\sqrt{1-\frac{M_{0}\ell ^{4}}{r^{4}}-\frac{%
\rho ^{2}\ell ^{4}}{6e^{2}r^{6}}}\,, \label{5D charged}
\end{equation}
which is just Eq.  (\ref{f_d,p}) with $p=2$ and $k=-1$. This is the only
way to have a PGB AdS black hole, as discussed in Sec. \ref{AAdS}.
The
black hole horizon equation (\ref{horizon(d,p)}) in this case can be reduced
to a polynomial of third order in the positive variable $x=\left( \frac{\rho
^{2}\ell ^{4}}{6e^{2}}\right) ^{-1/3}r_{+}^{2}$,
\begin{equation}
x^{3}-a\,x-1=0\,,\qquad a=\left( M_{0}+1\right)
\ell ^{4}\left( \frac{\rho ^{2}\ell ^{4}}{6e^{2}}\right) ^{-2/3}\,,
\end{equation}
where the real coefficient $a$ depends on the black hole
parameters. The polynomial can have three real roots in general. Analyzing
the roots as a function of the parameter $a$, we find that the horizon forms
for all values of $M_{0}$. Depending on the value of $a$, the polynomial can
have three zeros ($a>1,89$), two zeroes ($a=1,89$) or one zero ($a<0$), but
only one of them has a positive $x$ corresponding to a real horizon $r_{+}$,
as shown in Figure \ref{fig-horizon}. This implies that the black hole has only one
horizon for any $(M_0,\rho)$. We conclude that there are no naked singularities.
A black hole always forms, requiring $M_{0}\neq 0$ for the neutral objects.
The zero mass black hole exists only if $\rho \neq 0$.

Using the standard method, we calculate the Hawking temperature of the
obtained black holes,
\begin{equation}
T=\frac{1}{4\pi }\left( \frac{2r_{+}^{3}}{\ell ^{4}}+\frac{\rho ^{2}}{6e^{2}}%
\frac{1}{r_{+}^{3}}\right) \,.  \label{T}
\end{equation}
The temperature is finite and it behaves as $T\sim r_{+}^{3}$ for large
horizon radii. Note that the temperature of the PGB AdS black hole is very
different from that of the RN black hole, $T_{RN}\sim r_{+}$, which is one
of consequences of working with a higher-curvature theory. Another one is
that the entropy is not proportional to the horizon area, but to the horizon
line element.

The black hole temperature (\ref{T}) is never zero. Thus, there are no
extremal PGB AdS black holes.

\section{Pure Gauss-Bonnet gravity coupled to scalar field}

In this section we explore solutions of pure Gauss-Bonnet gravity
coupled to charged matter. We will focus on five-dimensional ($d=4$) bulk gravity for simplicity,
where $\Lambda =12/\ell^{4}$ follows from Eq.~(\ref{Eq-l}) and the
Gauss-Bonnet term ($p=2$) is the kinetic term for gravity. We are
interested in holographic applications, so we focus on asymptotically AdS
space and a gravitational field coupled to a so-called St\"{u}ckelberg holographic superconductor \cite{Franco:2009yz},
built from the Maxwell electromagnetic field, $A_{\mu }$, and the complex
St\"{u}ckelberg scalar, $\hat{\Psi}=\Psi e^{i\theta }$, where $\Psi (x)$
and $\theta (x)$ are real fields. Then the matter action is described by
\begin{equation}
I_{\text{M}}=-\frac{1}{2\kappa }\int d^{5}x\sqrt{-g}\left[ -\frac{1}{4e^{2}}
\,F^{2}-\frac{1}{2}\,(\partial \Psi )^{2}-\frac{m^{2}}{2}\,\Psi ^{2}
-\frac{1}{2}\,\mathcal{F}(\Psi )\left( \partial \theta -A\right) ^{2}\right] ,
\label{Eq-MatterAction}
\end{equation}
where a gauge invariant function $\mathcal{F}(\Psi )
=\Psi ^{2}+c_{3}\Psi^{3}+c_{4}\Psi ^{4}$ is positive to ensure positivity of the kinetic term
for the field $\theta $. The minimal coupling corresponds to $c_{3}=c_{4}=0$.
The total action of the system is
\begin{equation}
I=I_{0}+B=I_{\text{G}}+I_{\text{M}}+B\,,  \label{Eq-Itot}
\end{equation}
where $B$ is a boundary term to be added to the bulk term
$I_{0}=I_{\text{G}}+I_{\text{M}}$, so that the action principle for given boundary
conditions is satisfied.

\subsection{Equations of motion}

Variation of the total action leads to the gravitational equations of motion,
\begin{equation}
^{(2)}G_{\nu }^{\mu }=T_{\nu }^{\mu }\,.  \label{Eq-gravity}
\end{equation}
The generalized Einstein tensor
$^{(2)}G_{\nu }^{\mu }= H_{\nu }^{\mu }+\Lambda \,\delta _{\nu }^{\mu}$
includes, apart from the cosmological constant, the Lanczos tensor
\begin{eqnarray}
H_{\nu }^{\mu } &=&-\frac{1}{8}\,\delta _{\nu \nu _{1}\cdots \nu _{4}}^{\mu
\mu _{1}\cdots \mu _{4}}\,R_{\mu _{1}\mu _{2}}^{\nu _{1}\nu _{2}}R_{\mu
_{3}\mu _{4}}^{\nu _{3}\nu _{4}}\,,  \notag \\
&=&-\frac{1}{2}\,\delta _{\nu }^{\mu }\left( R^{2}-4R^{\alpha \beta
}R_{\alpha \beta }+R^{\alpha \beta \lambda \sigma }R_{\alpha \beta \lambda
\sigma }\right)  \notag \\
&&+2\left( RR_{\nu }^{\mu }-2R^{\mu \lambda }R_{\lambda \nu }-2R_{\lambda
\nu \sigma }^{\mu }R^{\lambda \sigma }+R^{\mu \alpha \lambda \sigma }R_{\nu
\alpha \lambda \sigma }\right) \,.
\end{eqnarray}
The energy-momentum tensor reads
\begin{eqnarray}
T_{\mu \nu } &=&\frac{1}{2e^{2}}\,\left( F_{\mu \lambda }F_{\nu }^{\ \lambda
}-\frac{1}{4}\,g_{\mu \nu }\,F^{2}\right) +\frac{1}{2}\,\partial _{\mu }\Psi
\partial _{\nu }\Psi +\frac{1}{2}\,\mathcal{F}(\Psi )\,\left( \partial _{\mu
}\theta -A_{\mu }\right) \left( \partial _{\nu }\theta -A_{\nu }\right)
\notag \\
&&-\frac{1}{4}\,g_{\mu \nu }\left[ (\partial \Psi )^{2}+m^{2}\Psi ^{2}+
\mathcal{F}(\Psi )\left( \partial \theta -A\right) ^{2}\right] \,.
\end{eqnarray}
The Maxwell and Klein-Gordon equations are, respectively,
\begin{eqnarray}
&&\nabla _{\nu }F^{\nu \mu }=-e^{2}\mathcal{F}(\Psi )\left( \nabla ^{\mu
}\theta -A^{\mu }\right) \,,  \notag  \label{Eq-matter} \\
&&\left( \square -m^{2}\right) \Psi =\frac{1}{2}\,\mathcal{F}^{\prime }(\Psi
)\,\left( \nabla \theta -A\right) ^{2}\,.
\end{eqnarray}
The field equation with respect to $\theta(x) $ is not independent due to a local
$U(1)$ symmetry. We will choose the gauge fixing $\theta =0$.

Consider a static, topological black hole metric in AdS space that
generalizes (\ref{static no hair}), keeping $g_{rr}$ and $g_{tt}$ as
independent functions due to the presence of scalar fields,
\begin{equation}
ds^{2}=g_{\mu \nu }\,dx^{\mu }dx^{\nu }=-f(r)\,dt^{2}+\frac{dr^{2}}{f(r)N(r)}
+r^{2}d\Omega ^{2}\,.  \label{Eq-BHmetric}
\end{equation}
Here, $d\Omega ^{2}=\gamma _{mn}(y)\,dy^{m}dy^{n}$ is a metric of the
transversal section of the unit radius and constant curvature $k$. The
boundary is placed at radial infinity, $r\rightarrow \infty $.

As we pointed out in Sec.~\ref{AAdS}, we will choose a hyperbolic horizon,
$k=-1$. One possible choice of the transversal coordinates is
\begin{equation}
d\Omega ^{2}=d\eta ^{2}+\sinh ^{2}\eta \,
\left( d\theta ^{2}+\sin ^{2}\theta\,d\varphi ^{2}\right),
\quad \eta\geq 0\,,\;\theta\in [0,\pi]\,, \; \phi\in [0,2\pi]\,.   \label{gamma}
\end{equation}
The horizon $r_{+}>0$ is the largest root of the equation $f(r_{+})=0$ such
that $N(r_{+})\neq 0$ and $f$, $N$ are positive functions when $r>r_{+}$. The black hole
temperature is calculated from
\begin{equation}
T=\frac{1}{4\pi }\,f^{\prime }(r_{+})\sqrt{N(r_{+})}\,.
\label{Eq-temperature}
\end{equation}

As usual, we assume that all fields possess the same isometries, that is, they are
static and spherically symmetric. Thus, the scalar field has the form
$\Psi =\Psi (r)$, and the gauge field,
$A_{\mu }=\phi \left( r\right) \,\delta _{\mu }^{t}$, generates
electric field $F_{tr}=-\phi ^{\prime }$ and $F^{2}=-2N\phi ^{\prime 2}$.
With this ansatz in hands, we can write the nonvanishing components of Eqs.~(\ref{Eq-gravity})
and (\ref{Eq-matter}) as
\begin{eqnarray}
0 &=&-\frac{6fN^{\prime }\left( fN-k\right) }{r^{3}}+\frac{1}{2}\,fN\,\Psi
^{\prime 2}+\frac{\mathcal{F}\phi ^{2}}{2f}\,,  \label{Eq-tt} \\
0 &=&-\frac{6Nf^{\prime }\left( fN-k\right) }{r^{3}}+\Lambda +\frac{N\phi
^{\prime 2}}{4e^{2}}-\frac{\mathcal{F}\phi ^{2}}{4f}+\frac{1}{4}\,m^{2}\Psi
^{2}-\frac{1}{4}\,fN\Psi ^{\prime }{}^{2}\,,  \label{Eq-rr} \\
0 &=&\frac{3Nf\,\Psi ^{\prime }}{r}+\frac{N^{\prime }f\,\Psi ^{\prime }}{2}%
+Nf^{\prime }\,\Psi ^{\prime }+Nf\,\Psi ^{\prime \prime }-m^{2}\Psi +\frac{%
\phi ^{2}\mathcal{F}^{\prime }}{2f}\,,  \label{Eq-scalar} \\
0 &=&\frac{3N\phi ^{\prime }}{r}+\frac{\phi ^{\prime }N^{\prime }}{2}+N\phi
^{\prime \prime }-\frac{e^{2}\mathcal{F}\phi }{f}\,.  \label{Eq-Maxwell}
\end{eqnarray}%
Independent gravitational equations are along $rr$ and $tt$. Note that
Eq.~(\ref{Eq-tt}) is a difference of the original equations that arises from
a backreaction of the gravitational field, leading to $g_{rr}g_{tt}\neq -1$
($N\neq 1$) and in general different $rr$ and $tt$ energy-momentum components,
\begin{equation}
T_{t}^{t}-T_{r}^{r}=-\frac{1}{2}\,fN\,\Psi ^{\prime 2}-\frac{\mathcal{F}\phi
^{2}}{2f}\,.  \label{Ttt}
\end{equation}%
Without matter fields, $T^t_t = T^r_r $ and Eq.(\ref{Eq-tt}) uniquely gives $N(r) = 1$. When the matter is present, the energy-momentum tensor has to satisfy the weak energy condition $%
T_{\mu \nu }\,u^{\mu }u^{\nu }\leq 0$ for the timelike unit vector $u^{\mu
} $. Explicitly, in our ansatz with $(u_{t},u_{i})=(-\sqrt{f},0)$, it reads
\begin{equation}
T_{t}^{t}=\frac{1}{4e^{2}}\,N\phi ^{\prime 2}+\frac{1}{4}\,m^{2}\Psi ^{2}+%
\frac{1}{4}\,fN\Psi ^{\prime 2}+\frac{\mathcal{F}\phi ^{2}}{4f}\geq 0\,.
\label{WEcondition}
\end{equation}%
Without the scalar field, the condition is fulfilled since $T_{t}^{t}=\frac{1%
}{4e^{2}}\,\phi ^{\prime 2}\geq 0$. With the scalar field, the functions $N$%
, $f$, $\mathcal{F}$ are positive outside the black hole, so the only term
that can decrease the energy density $T_{t}^{t}$ is due to the negative mass
of the scalar field. If black hole hair is short,
the scalar field decays fast as it goes to the boundary and is nontrivial
close to the horizon. Let us analyze, therefore, the weak energy condition
on the horizon. Using the gravitational equation (\ref{Eq-rr}), we obtain
\begin{equation}
T_{t}^{t}(r_{+})=24\pi T\,\frac{\sqrt{N(r_{+})}}{r_{+}^{3}}-\frac{12}{\ell ^{4}}
\geq 0\,.
\end{equation}

Our next step is to solve the above equations and determine the unknown
functions $f(r)$, $N(r)$, $\phi (r)$ and $\Psi (r)$.

\subsection{Boundary conditions for the fields \label{Asymptotic}}

The equations of motion are not exactly solvable in a given ansatz when the scalar field is nonvanishing.
To solve a system of second order differential equations, we have to
specify the behavior of the fields at the horizon, $r=r_{+}$, and at the
asymptotic boundary, $r\rightarrow \infty $.\bigskip

\textbf{i) Behavior at the horizon} ($r=r_{+}$)\medskip

We already discussed the behavior of the gravitational fields, which is
\begin{eqnarray}
f(r_{+}) &=&0\,,\qquad f^{\prime }(r_{+})=\dfrac{4\pi T}{\sqrt{N(r_{+})}}=\
\text{finite}\,,  \notag \\
N(r_{+}) &=&\,\text{finite}\neq 0\,.
\end{eqnarray}%
For the electric potential, we choose the boundary condition
\begin{equation}
\phi (r_{+})=0\,,
\end{equation}%
which ensures a finite effective mass of the scalar field in the probe
limit.

The scalar field has to be finite on the horizon. Using the equation of
motion for the scalar field, we get%
\begin{equation}
\Psi ^{\prime }(r_{+})=\dfrac{m^{2}}{4\pi T}\,\frac{\Psi (r_{+})}{\sqrt{%
N(r_{+})}}=\text{ finite}\,.
\end{equation}%

\textbf{ii) Behavior at the boundary} ($r\rightarrow \infty $)\medskip

The gravitational field (\ref{Eq-BHmetric})\ must be asymptotically locally
AdS, so that we impose
\begin{eqnarray}
f &\simeq &k+\dfrac{r^{2}}{\ell ^{2}}-\dfrac{M_{0}\ell ^{2}}{2r^{2}}\,,
\notag \\
fN &\simeq &k+\dfrac{r^{2}}{\ell ^{2}}-\dfrac{M\ell ^{2}}{2r^{2}}\,,  \notag
\\
N &=&\frac{fN}{f}\simeq 1+\dfrac{\left( M_{0}-M\right) \ell ^{4}}{2r^{4}}\,.
\label{fN_asympt}
\end{eqnarray}%
The mass parameters are normalized so that, without the scalar field, they
reduce to the known result of Sec. \ref{PGB AdS black hole} with $M=M_{0}$.
The black hole mass is related to the parameter $M$. The scalar hair,
thus modifies the mass of the black hole through the function $N\neq 1$.

Related to the electric potential, the quantity of physical interest is the
chemical potential $\mu =\phi (\infty )-\phi (r_{+})=\phi (\infty )$ which
represents the potential at infinity measured with respect to the event
horizon. Based on the Maxwell equation, the electric potential behaves
asymptotically as%
\begin{equation}
\phi (r)\simeq \mu -\frac{\rho }{r^{2}}\,.  \label{phi_asympt}
\end{equation}

Finally, the scalar field must be finite everywhere so that it can be
interpreted as the black hole hair. From the scalar equation (\ref{Eq-scalar}),
in the asymptotically AdS sector (\ref{fN_asympt}) and (\ref{phi_asympt}), we get%
\begin{equation}
0\simeq \Psi ^{\prime \prime }+\frac{5}{r}\,\Psi ^{\prime }-\frac{m^{2}\ell
^{2}}{r^{2}}\,\Psi \,,
\end{equation}%
leading to the asymptotic solution%
\begin{equation}
\Psi (r)\simeq \frac{\Psi _{-}}{r^{\Delta _{-}}}+\frac{\Psi _{+}}{r^{\Delta
_{+}}}\,.  \label{Eq-BehaInf}
\end{equation}%
Here, $\Delta _{\pm }=2\pm \sqrt{4+m^{2}\ell ^{2}}$ is a conformal dimension
of the scalar operators $\Psi _{\pm }=\left\langle \mathcal{O}_{\pm
}\right\rangle $ in a dual conformal field theory (CFT).

A black hole with scalar hair forms when the background decreases the
effective mass of the scalar field so that it becomes negative. This causes a
breaking of the $U(1)$ gauge symmetry in the bulk gravity, which is dual to
a condensation operator $\mathcal{O}$ breaking the global $U(1)$ symmetry at
the boundary \cite{Gubser:2008wv}.

A choice of the boundary conditions for the scalar field determines which
quantity will be kept fixed on the boundary. If it is $\Psi _{-}$ (Dirichlet
boundary conditions), then the term $\Psi _{-}$ becomes a source in the
holographic quantum field theory (QFT) and a vacuum expectation value (VEV) of a scalar operator
$\mathcal{O}_{+}$ of conformal dimension $\Delta _{+}$ is identified with the bulk
operator, $\Psi _{+}=\left\langle \mathcal{O}_{+}\right\rangle $. It is also
possible to keep $\Psi _{+}$ fixed on the boundary (Neumann boundary
conditions), when $\Psi _{+}$ becomes a source and $\Psi _{-}=\left\langle
\mathcal{O}_{-}\right\rangle $ is the VEV.

On the other hand, the scalar field is finite when $\Delta _{\pm }\geq 0$,
which implies that the dual operator is relevant or marginal and can be
switched on without destroying the UV fixed point in CFT. The dual CFT
is unitary for the masses of the scalar field that take values in the
Breitenlohner-Freedman window, $-4\leq m^{2}\ell ^{2}\leq -3$, or
equivalently $1\leq \Delta _{-}\leq 2$, or $2\leq \Delta _{+}\leq 3$. The
conformal anomaly is absent if $\Delta _{\pm }\neq 4$.

In the following discussion, we will set $m^{2}\ell ^{2}=-3$, which corresponds to $\Delta _{-}=1$ and $\Delta _{+}=3$, and choose the Neumann
boundary condition$\ $where $\Psi _{-}=0$. Then the response operator is $%
\Psi _{+}=\left\langle \mathcal{O}_{+}\right\rangle $ so that
\begin{equation}
\Psi (r)\simeq \frac{\left\langle \mathcal{O}_{+}\right\rangle }{r^{3}}\,.
\end{equation}

\section{Black hole thermodynamics}

Thermal properties of the black hole can be obtained from the partition
function evaluated in semiclassical approximation,%
\begin{equation}
Z=e^{-I_{\text{class}}^{E}}\,,
\end{equation}%
where $I_{\text{class}}^{E}$ is the classical Euclidean action. In an
asymptotically AdS spacetime this action is divergent and should be
renormalized. Its finite part contains the thermodynamic information
about the system through the quantum statistical relation
\begin{equation}
TI^{E}=U+\mu Q-TS\,,  \label{QSR}
\end{equation}
where $U$ is the total internal energy of the system at the temperature $T$,
the constant $Q$ is its electric charge and $S$ is the black hole entropy. In the
framework of AdS/CFT correspondence, the Euclidean gravity action is
identified with the thermodynamic potential (free energy) $G=TI^{E}$ of the holographic QFT. Also,
the asymptotic charges in AdS space are interpreted as the thermodynamic
charges in a boundary QFT. Finally, in the holographic dictionary, the black hole temperature and entropy
match the field theory temperature and entropy, respectively.

Thus, in order to obtain the finite quantities in asymptotically AdS gravity (IR finiteness) and
holographic QFT (UV finiteness), we need a renormalized gravitational action.

\subsection{Renormalized action and Gibbs free energy}

Euclidean spacetime $(\tau, r, y^m)$ is obtained from the Lorentzian spacetime $(t,r,y^m)$
 by performing the Wick rotation of the
temporal coordinate ($t=i\tau $) with the period of the Euclidean time $T^{-1}$
which avoids the conical singularity at the horizon. The Euclidean
on-shell action $I^{E}=-iI$ has the form
\begin{equation}
I^{E}=\frac{V_{3}}{T}\int\limits_{r_{+}}^{\infty }dr\,\frac{r^{3}}{%
\sqrt{N}}\,\mathcal{L}(r)\,. \label{Euclid}
\end{equation}
The volume $V_{3}=\int d^{3}y\,\sqrt{\gamma }=\int \sinh ^{2}{\eta }\sin {%
\theta }d\eta d\theta d\phi $ of the hyperbolic transversal section is
infinite, so all physical quantities are taken per unit volume.

Let us first evaluate the Euclidean bulk action, $I_{0}$, which is a sum of
the gravitational part (\ref{PGB}) and the matter part (\ref{Eq-MatterAction}).
Substituting the equations of motion, all explicit contributions of the scalar field
$\Psi$ cancel out, so that the information about the black hole hair is contained in the function $N\neq 1$.
Furthermore, the Euclidean on-shell action can be written as a total derivative,
\begin{eqnarray}
I_{0}^{E} &=&\frac{V_{3}}{2\kappa T}\int\limits_{r_{+}}^{\infty }dr\,\left[
12r\sqrt{N}\,f^{\prime }\left( fN-k\right) +\frac{r^{3}}{e^{2}}\,\sqrt{N}%
\phi \phi ^{\prime }\right] ^{\prime }  \notag  \label{Eq-EucliI0} \\
&=&\frac{V_{3}}{2\kappa T}\left. \left[ 12r\sqrt{N}\,f^{\prime }\left(
fN-k\right) +\frac{r^{3}}{e^{2}}\,\sqrt{N}\phi \phi ^{\prime }\right]
\right\vert _{r_{+}}^{\infty }\,.\label{I0E}
\end{eqnarray}%
Using the boundary conditions given in Sec. \ref{Asymptotic}, one can see that $%
I_{0}^{E}$ is divergent at infinity. We have not checked yet whether the
action is stationary on-shell for the chosen boundary conditions.

Thus, we search for the suitable boundary terms for the gravitational ($B_{%
\text{G}}$) and matter ($B_{\text{M}}$) fields,%
\begin{equation}\label{B}
B=B_{\text{G}}+B_{\text{M}}\,,
\end{equation}%
so that the total action is convergent and that it has a well-posed action
principle. We choose the Gauss-normal
frame ($g_{ri}=0$) in the local coordinates $x^{\mu }=(r,x^{i})$ where the
boundary is placed at the constant radius, $r=r_{B}\rightarrow \infty $.
Relevant quantities to describe the boundary dynamics are the induced metric, $h_{ij}$, and the extrinsic
curvature, $K_{ij}=-h_{ij}^{\prime }/2\sqrt{g_{rr}}$.

In a higher-order curvature AdS gravity, the simplest way to renormalize
the gravitational action $I_{\text{G}}$ is to add a unique boundary term, the 
 so-called Kounterterm \cite{Olea:2006vd,Kofinas:2008ub}, which
depends explicitly on the extrinsic curvature. Its form in
Einstein-Gauss-Bonnet AdS theory is given in Refs.~\cite{Olea:2006vd,Kofinas:2006hr} and the PGB Kounterterm is obtained by
leaving only the terms with the Gauss-Bonnet coupling,
\begin{equation}
B_{\text{G}}=\frac{3}{4\kappa }\int\limits_{r_{B}}d^{4}x\,\sqrt{-h}\,\delta
_{i_{1}\cdots i_{4}}^{j_{1}\cdots j_{4}}\,K_{j_{1}}^{i_{1}}\delta
_{j_{2}}^{i_{2}}\left( \mathcal{R}%
_{j_{3}j_{4}}^{i_{3}i_{4}}-K_{j_{3}}^{i_{3}}K_{j_{4}}^{i_{4}}+\frac{1}{3\ell
^{2}}\,\delta _{j_{3}}^{i_{3}}\delta _{j_{4}}^{i_{4}}\right) \,,  \label{Kt}
\end{equation}%
where $\mathcal{R}_{\ jkl}^{i}(h)$ is the intrinsic curvature of the
boundary. For more details on the method, see Refs.~\cite{Kofinas:2008ub,Miskovic:2007mg,Liu:2008zf}.

It is worthwhile noticing that the approach based on Kounterterms is
 equivalent to the standard one \cite{Myers:1987yn,Emparan:1999pm}
where the action principle based on Dirichlet boundary conditions for the
boundary metric $h_{ij}$ requires the PGB action to be supplemented by a
generalized Gibbons-Hawking term,%
\begin{equation}
B_{\text{GGH}}=\frac{2}{\kappa }\int\limits_{r_{B}}d^{4}x\,\sqrt{-h}\,\delta
_{i_{1}i_{2}i_{3}}^{j_{1}j_{2}j_{3}}\,K_{j_{1}}^{i_{1}}\,\left( \frac{1}{2}\,%
\mathcal{R}_{j_{2}j_{3}}^{i_{2}i_{3}}-\frac{1}{3}%
\,K_{j_{2}}^{i_{2}}K_{j_{3}}^{i_{3}}\right) \,.
\end{equation}
Indeed, the Dirichlet action $I_{\text{G}}+I_{\text{GGH}}$ has a variation
proportional to $\delta h_{ij}$,
\begin{equation}
\delta (I_{\text{G}}+B_{\text{GGH}}) =\frac{1}{\kappa }
\int\limits_{r_{B}}d^{4}x\,\sqrt{-h}\,\delta
_{i\,i_{1}i_{2}i_{3}}^{j\,j_{1}j_{2}j_{3}}\,(h^{-1}\delta
h)_{j}^{i}\,K_{j_{1}}^{i_{1}}\left( \frac{1}{2}\,\mathcal{R}%
_{j_{2}j_{3}}^{i_{2}i_{3}}-\frac{1}{3}\,K_{j_{2}}^{i_{2}}K_{j_{3}}^{i_{3}}
\right),
\end{equation}
because all variations $\delta K^{i}_{j}$ cancel out.
 In consequence, $I_{\text{G}}+B_{\text{GGH}}$ vanishes for the
Dirichlet boundary condition on the induced metric.
However, this action is  divergent
and one has to add the counterterms that cancel the divergences and do not change the Dirichlet boundary condition
(i.e., which depend only on the intrinsic quantities),
\begin{equation}
B_{\text{ct}}=\frac{1}{\ell \kappa }\int\limits_{r_{B}}d^{4}x\sqrt{-h}
\left( \frac{4}{\ell ^{2}}-\mathcal{R}\right) \,.
\end{equation}
It can be shown in a near-boundary analysis that the surface terms
$B_{\text{GGH}}+B_{\text{ct}}$ are equivalent to the Kounterterm
$B_{\text{G}}$ given by (\ref{Kt}). We shall use the last one, as it is
simpler and can be generalized to any dimension.
(While the form of $B_{\text{GGH}}$ is known in any $d$, the full
series for $B_{\text{ct}}$ is still unknown.) Evaluated on-shell, it becomes
\begin{equation}
B_{\text{G}}^{E}=\frac{9V_{3}}{2\kappa T}\,\lim_{r\rightarrow \infty }\sqrt{N%
}\left[ f\left( \frac{r^{2}}{\ell ^{2}}+2k-fN\right) +\frac{rf^{\prime }}{2}%
\left( 2k+\frac{r^{2}}{3\ell ^{2}}-3fN\right) \right] ,
\end{equation}
and  in asymptotically AdS spaces it has the form
\begin{equation}\label{BGE}
B_{\text{G}}^{E}=-\frac{6V_{3}}{\kappa T}\,\lim_{r\rightarrow \infty }\left(
\frac{2r^{4}}{\ell ^{4}}-\frac{3k^{2}}{4}-2M+\frac{3M_{0}}{2}\right) \,.
\end{equation}
Then combining equations (\ref{I0E}), (\ref{B}) and (\ref{BGE}), we have the renormalized action
\begin{equation}
I^{E}=\frac{V_{3}}{\kappa T}\,\left( \frac{9}{2}+3M\right) +
\frac{V_{3}\mu \rho }{\kappa Te^{2}}-\frac{24\pi V_{3}}{\kappa }\,r_{+}
+B_{\text{M}}^{E}\,.  \label{Iren}
\end{equation}

The matter fields do not contain IR divergences in the chosen
coupling, but we can need the surface term $B_{\text{M}}$ to ensure the
stationary on-shell action.
\medskip

Thus, let us check the variational principle. We already clarified that the
gravitational action is based on Dirichlet boundary conditions for the
induced metric. When we vary the matter field and use the equations of motion, we obtain
\begin{equation}
-\frac{1}{2\kappa }\int\limits_{r_{B}}d^{4}x\sqrt{-h}\, n_r \left( F_{rt}\,\delta
\phi -g^{rr}\Psi'\delta \Psi \right) \,,  \label{Eq-boundarymatter}
\end{equation}
where $n_\mu=(n_r,n_i)=(\sqrt{f},0)$ is the unit normal to the asymptotic boundary.
The above expression must vanish or be canceled out by the boundary term $B_{\text{M}}$.
For the electromagnetic field, we can choose a grand canonical ensemble where
the chemical potential is kept fixed on the boundary, $\delta \phi =0$,
which will make the first term in (\ref{Eq-boundarymatter}) vanish.
Then well-defined variation principle for the Maxwell field
does not require any surface term. Alternatively, if one considers
a canonical ensemble where the  charge density is kept fixed, $\delta \phi ^{\prime
}=0$,  a new term,  $\frac{1}{2\kappa e^{2}}\int_{r_{B}}d^{4}x\sqrt{-h}\,n_r\,\phi \,F^{tr}$ has to be added.

Similarly, there are at least two possible choices of a boundary condition for the scalar field:
the Dirichlet condition ($\delta \Psi =0$) and the Neumann one ($\delta \Psi'=0$).
In the former case, one does not need new boundary terms in the action to have its
variation well defined, while in the latter case one should add
$-\frac{1}{2\kappa }\int_{r_{B}}d^{4}x\sqrt{-h}\,n^r\,\Psi \Psi'$. Holographically,
these two choices correspond to different quantizations in the quantum field theory,
with $\Psi_{-}=0 $ or $\Psi _{+}=0$ in Eq.~(\ref{Eq-BehaInf}), identifying the scalar
field with the VEV or the source, respectively.

In this work, the mass of the scalar field is set to $m^{2}\ell
^{2}=-3$. We shall also choose the Neumann boundary conditions for the scalar
field, making the source vanish, $\Psi _{-}=0$.
With respect to the 
electromagnetic field, we shall work in a grand canonical ensemble by
fixing the chemical potential. Thus, the boundary term for the matter fields
reads
\begin{equation}
B_{\text{M}}=-\frac{1}{2\kappa }\int\limits_{r_{B}}d^{4}x\sqrt{-h}\, n^r\,\Psi \Psi'\,. \label{Bm}
\end{equation}
Evaluating it on-shell we find that, due to a fast falloff of $\Psi(r)$, the scalar field boundary term
does not contribute, $B_{\text{M}}^{E}=0$.

In order to interpret the expression for the free energy (\ref{Iren}), we
have to calculate the conserved charges and entropy of the system.

\subsection{Conserved charges \label{Cc}}

Total internal energy of the black hole can be obtained as the Noether
charge associated with the asymptotic Killing vector $\xi^i =\delta _{t}^{i}$
for time translations, evaluated at the transversal section of spacetime, $t$,
$r=Const$, denoted by $\Sigma _{r}$, through the formula \cite{Olea:2006vd}
\begin{equation}
U=\int\limits_{\Sigma _{\infty }}d^{3}y\sqrt{\sigma }u_{j}\xi ^{i}\left(
q_{(0)i}^{j}+q_{i}^{j}\right) \,.
\end{equation}%
Here, $u_{i}=-\sqrt{-g_{tt}}\,\delta _{i}^{t}$ is the unit normal to the
surface $\Sigma _{r}$ described by the transversal metric $\sigma
_{mn}=r^{2}\gamma _{mn}$ and $\gamma _{mn}$ is the hyperbolic metric given
by Eq.(\ref{gamma}). It is well known that in five dimensions the total
gravitational energy includes the vacuum energy ($q_{(0)i}^{j}$ term), black
hole mass $M_{\text{BH}}$ (gravitational $q_{i}^{j}$ contribution) and
energy of the matter fields (matter $q_{i}^{j}$ contribution), that is,%
\begin{equation}
U=E_{\text{vac}}+M_{\text{BH}}+M_{\text{M}}\,.
\end{equation}%
The vacuum energy exists in odd bulk dimensions only, and it corresponds to
the energy of the empty (global) AdS space. In particular, the vacuum energy
in Einstein-Gauss-Bonnet (EGB) gravity with topological black holes is given in Ref.~\cite{Kofinas:2006hr},
from where we can deduce the five-dimensional PGB
expression by keeping only the Gauss-Bonnet contribution,
\begin{equation}
E_{\text{vac}}=\int\limits_{\Sigma _{\infty }}d^{3}y\sqrt{\sigma }%
u_{t}\,q_{(0)t}^{t}=\frac{V_{3}}{\kappa }\,\frac{9k^{2}}{2}\,.  \label{Evac}
\end{equation}
On the other hand, it was shown in Refs.~\cite{Jatkar:2014npa,Jatkar:2015ffa} that the
gravitational part of the charge density tensor $q_{i}^{j}$ in EGB gravity
can be written in terms of the Weyl tensor, $W_{\alpha \beta }^{\mu \nu
}=R_{\alpha \beta }^{\mu \nu }-\frac{4}{3}\,\delta _{(\alpha }^{(\mu
}R_{\beta) }^{\nu) } +\frac{1}{12}\,\delta _{\alpha \beta}^{\mu \nu} R$ or, more precisely, its
electric part $W_{j\nu }^{i\mu }\,n_{\mu }n^{\nu }=W_{jr}^{ir}$ as
\begin{equation}
M_{\text{BH}}=\frac{1}{2\kappa \ell }\int\limits_{\Sigma _{\infty }}d^{3}y
\sqrt{\sigma }u^{j}\xi ^{i}\,W_{jr}^{ir}\,,
\end{equation}
where $n_{\mu }=\delta _{\mu }^{r}\left( fN\right) ^{-1/2}$ is the normal
vector to the spacetime boundary $r=Const$. In fact, since the Weyl tensor is
traceless, the quantity that enters the black hole mass is
\begin{equation}
W_{jr}^{ir}=-W_{jk}^{ik}=-\left( R_{jk}^{ik}+\frac{1}{\ell ^{2}}\,\delta
_{jk}^{ik}\right) +\mathcal{O}(1/r^{8})\,,
\end{equation}
so that it can be easily evaluated with the help of (\ref{fN_asympt}) as
\begin{equation}
M_{\text{BH}}=\frac{6}{\kappa \ell }\,\int\limits_{\Sigma _{\infty
}}d^{3}y\,r^{3}\sqrt{\gamma }\sqrt{f}\,\left( \frac{fN-k}{r^{2}}-\frac{1}{%
\ell ^{2}}\right) =\frac{V_{3}}{\kappa }\,3M\,.
\end{equation}
We can also show that $M_{\text{M}}=0$, so the total internal energy
of the black hole (the vacuum energy plus its mass) is
\begin{equation}
U=\frac{V_{3}}{\kappa }\,\left( \frac{9k^{2}}{2}+3M\right) \,.
\label{U}
\end{equation}

The action is also invariant under the local $U(1)$ transformations $\delta
_{\lambda }A_{\mu }=\partial _{\mu }\lambda $, $\delta _{\lambda }\theta
=\lambda $ ($\psi $, $g_{\mu \nu }$ do not transform). The Noether charge is calculated
 from the electromagnetic current $J^{\mu
}(\lambda )=\frac{1}{2\kappa e^{2}}\,\partial _{\nu }\left( \lambda \sqrt{-g}%
F^{\mu \nu }\right) $ with $\lambda=1$,
\begin{equation}
Q=\int\limits_{r_{B}}d^{4}x\,J^{r}=\frac{V_{3}}{2\kappa }\,\lim\limits_{r%
\rightarrow \infty }\left( r^{3}\sqrt{N}\,\phi ^{\prime }\right) =\frac{%
V_{3}\rho }{\kappa e^{2}}\,.  \label{Qe}
\end{equation}

Found Noether charges $U$ and $Q$ should match the thermodynamic charges
\begin{eqnarray}
U &=&G-T\left( \frac{\partial G}{\partial T}\right) _{\mu }
-\mu \left( \frac{\partial G}{\partial \mu }\right) _{T}\,,  \nonumber \\
Q &=&\left( \frac{\partial G}{\partial \mu }\right) _{T}\,,
\end{eqnarray}
obtained from the thermodynamic partition function using the first law of
thermodynamics. The thermodynamic charges are the ones that enter the quantum
statistical relation.

\subsection{Quantum statistical relation}

To describe a thermodynamic system, we have to know its entropy. The
entropy of Lovelock AdS gravities is given, for example, in
Ref.~\cite{Kofinas:2007ns}. Applying this formula for static, spherically symmetric, topological black holes
in pure Lovelock gravity with the coupling constant $\alpha _{p}$, we obtain
\begin{equation}
S=\frac{\left( d-1\right) !V_{d-1}p\alpha _{p}}{4G\left( d-2p+1\right) !}%
\,r_{+}^{d-2p+1}k^{p-1}\,. \label{Sp}
\end{equation}
In the PGB case ($p=2$) in five dimensions ($d=4$) with
$16\pi G=2\kappa $ and $\alpha _{2}=-1$, and with the hyperbolic horizon ($%
k=-1)$, the above expression becomes
\begin{equation}
S=\frac{24\pi V_{3}}{\kappa }\,r_{+}\,. \label{S}
\end{equation}
It is worthwhile noticing that $S$ is positive only for the hyperbolic
horizons, which are the only black holes that exist in the PGB gravity in
AdS space.

Now we can interpret Eq.(\ref{Iren}) in the grand canonical ensemble
($B_{\text{M}}^{E}=0$) as the quantum statistical relation of the system (\ref{QSR}).
Indeed, replacing the expressions for the total energy (\ref{U}),
the electric charge (\ref{Qe}) and the entropy (\ref{S}),  we obtain
\begin{equation}
G=TI^{E}=U+\mu Q-TS\,,
\end{equation}
which is nothing but the Legendre transformation of the Gibbs potential $G$.
Then the first law of thermodynamics, $\delta U=T\delta
S-\mu \delta Q$, can be equivalently written as $\delta G=Q\delta \mu -S\delta T$.

Let us discuss the case $\Psi =0$ where the exact solution is known. From $\phi (r_{+})=0$ and
$f(r_{+})=0$, we calculate the charges,
\begin{eqnarray}
M_{0} &=&\frac{r_{+}^{4}}{\ell ^{4}}-1-\frac{\mu ^{2}r_{+}^{2}}{6e^{2}}\,,
\nonumber \\
Q &=&\frac{V_{3}\mu r_{+}^{2}}{\kappa e^{2}}\,,
\end{eqnarray}
and Eq.(\ref{T}) gives the temperature
\begin{equation}
T=\frac{1}{4\pi }\left( \frac{2r_{+}^{3}}{\ell ^{4}}
+\frac{\mu ^{2}r_{+}}{6e^{2}}\right) \,.
\end{equation}
Using the expression (\ref{U}) for the internal energy, we obtain the
free energy in terms of $r_{+}$ and $\mu $ [because its natural
variables are $T(r_{+},\mu )$ and $\mu $],
\begin{equation}
G=\frac{V_{3}}{\kappa }\,\left( \frac{3}{2}-\frac{9r_{+}^{4}}{\ell ^{4}}-
\frac{\mu ^{2}r_{+}^{2}}{2e^{2}}\right) \,. \label{Gnormal}
\end{equation}
It is straightforward to show by varying $G$ in $r_{+}$ and $\mu $ that the
first law of thermodynamics is fulfilled.

In the next section we perform a similar calculation for the case $\Psi \neq 0$.

\section{Holographic phase transition}

Having an asymptotically AdS space and the scalar field turned on, we can
use the AdS/CFT correspondence tools to study a dual quantum theory. A
particular black hole solution breaks the conformal symmetry on the boundary
and leads to a holographic theory which is thermal. Our goal is to analyze the
possibility of having a phase transition in the four-dimensional QFT due to a
change of temperature. In practice, this means that we have to find a
backreaction solution of the system. Since it is not exactly solvable, we
shall integrate numerically a set of equations (\ref{Eq-tt})-(\ref{Eq-Maxwell})
and use the probe limit to simplify it. Namely, in this limit the gravity dynamically decouples from
the matter, and the scalar field moves in the black hole background.

In addition, when the mass of the scalar field saturates the upper Breitenlohner-Freedman bound, the gravitational
backreaction could modify the asymptotic behavior of the theory, and the free energy in asymptotically AdS space
would require additional surface terms in order to become regular \cite{Henneaux:2006hk}. Thus, the dynamics without
the gravitational backreaction better catches a typical behavior of the system.

For numerical calculations it is convenient to introduce a new dimensionless
variable, $z=r_{+}/r$. All functions are defined in the region $z\in (0,1]$,
where $z=1$ is the location of the horizon and $z=0$ corresponds to the
asymptotic boundary. We can set $r_{+}=1$ by the following rescaling,
\begin{equation}
\begin{array}[b]{llllll}
r & \rightarrow r_{+}r\,,\qquad & \Psi & \rightarrow r_{+}\Psi \,,\qquad &
\ell & \rightarrow \ell /r_{+}\,, \\
f & \rightarrow r_{+}^{2}\,f\,, & \phi & \rightarrow r_{+}^{2}\phi \,, & m &
\rightarrow r_{+}m\,, \\
N & \rightarrow r_{+}^{2}N\,, & \mathcal{F} & \rightarrow r_{+}^{2}\mathcal{F
}, & k & \rightarrow r_{+}^{4}\,k\,.
\end{array}
\label{scaling}
\end{equation}

The probe approximation is obtained as the large charge limit, $e\rightarrow
\infty $, after rescaling $\Psi =\frac{1}{e}\tilde{\Psi}$ in the Eqs.
(\ref{Eq-tt})-(\ref{Eq-Maxwell}) \cite{Franco:2009yz}.
The scaling properties of the function $\mathcal{F}\left( \Psi \right) $ are
determined from the behavior of $\Psi ^{n}$, whose dimensional analysis
gives $\left. \mathcal{F}\sim c_n (eL)^{n-2}\Psi^{n}\right.$,
where $L$ is some length scale and $c_n$ a dimensionless constant,
implying that $\mathcal{F}( \frac{1}{e}\tilde{\Psi}) \rightarrow 0$ when $e\rightarrow \infty $. In general,
we shall require that
\begin{eqnarray}
 \mathcal{\tilde{F}} ( \tilde{\Psi} ) &=& \lim_{e\rightarrow \infty }
\left[ e^{2}\mathcal{F}\left( \frac{1}{e}\tilde{\Psi}\right) \right]
<\infty \,,  \notag \\
\mathcal{\tilde{F}}' (\tilde{\Psi} )
&=&\lim_{e\rightarrow \infty }\left[ e^{2}\tfrac{d}{d\tilde{\Psi}}\mathcal{F}
\left( \frac{1}{e}\tilde{\Psi}\right) \right] <\infty \,\,.
\end{eqnarray}%
Then the gravitational equations (\ref{Eq-tt}) and (\ref{Eq-rr}) solve $N=1$
and (\ref{Eq-rr})-(\ref{Eq-Maxwell}) in the probe limit become
\begin{eqnarray}
0 &=&\frac{6f^{\prime }\left( f-k\right) }{r^{3}}-\Lambda \,,  \notag \\
0 &=&\frac{\left( r^{3}f\,\Psi ^{\prime }\right) ^{\prime }}{r^{3}}
-m^{2}\Psi +\frac{\phi ^{2}\mathcal{F}^{\prime }}{2f}\,,  \notag \\
0 &=&\frac{\left( r^{3}\phi ^{\prime }\right) ^{\prime }}{r^{3}}-\frac{%
\mathcal{F}\phi }{f}\,,  \label{probe limit}
\end{eqnarray}%
where we drop tildes for the sake of simplicity. The gravitational PGB
solution is a neutral hyperbolic black hole with the metric function
\begin{equation}
f(r)=-1+\frac{r^{2}}{\ell ^{2}}\sqrt{1-\frac{M_{0}\ell ^{4}}{r^{4}}}\,,
\end{equation}%
which is just Eq. (\ref{5D charged}) with the electric charge switched off. The
boundary conditions are the same as the ones presented in Sec. \ref{Asymptotic},
where now $N=1$ and thus $M=M_{0}$.

Since the gravity part has been decoupled, we have to focus only on the
matter action  which, in the black hole background, has the form
\begin{equation}
I_{\text{M}}=\frac{1}{4\kappa e^{2}}\,\int d^{5}x\sqrt{\gamma }\,r^{3}\left(
-\phi ^{\prime 2}+f\Psi ^{\prime 2}+m^{2}\Psi ^{2}-\frac{1}{f}\,\phi ^{2}%
\mathcal{F}\right) \,.
\end{equation}%
In the grand canonical ensemble and with the scalar field satisfying the Neumann boundary conditions,
we have to add the matter boundary term (\ref{Bm}).
Using the equations of motion, the Euclidean continuation of $I_{\text{M}}+B_{\text{M}}$
is
\begin{equation}
I_{0}^{E}+B_{\text{M}}^{E}=\frac{V_{3}}{4\kappa e^{2}T}\left[ \left. \left(
\rule{0pt}{14pt}r^{3}f\Psi \Psi ^{\prime }+r^{3}\phi \phi ^{\prime }\right)
\right\vert _{r_{+}}^{\infty }-\int_{r_{+}}^{\infty }dr\,\frac{r^{3}\phi
^{2}\Psi \mathcal{F}^{\prime }}{2f}\right] \,.
\end{equation}
Compared to the system that includes the backreaction, the on-shell
action is not a total derivative, and the nonlocal term has to be
evaluated numerically between the horizon and the asymptotic boundary.
Another important difference with respect to the backreaction noted in Ref.~\cite{Aranguiz:2013cwa},
is that the action needs scalar field counterterms when evaluated in the
probe limit, since it becomes IR divergent. This counterterm is discussed in
Ref.~\cite{Franco:2009yz} and it does not contribute to the result for our boundary conditions, as we
are allowed to set $\Psi _{-}=0$.

The finite free energy, $G=T(I_{0}^{E}+B_{\text{M}}^{E})$, has the form
\begin{equation}
G=\frac{V_{3}}{2\kappa e^{2}}\left( \rho \mu -\int_{r_{+}}^{\infty }dr\,
\frac{r^{3}\phi ^{2}\Psi \mathcal{F}^{\prime }}{4f}\right) .
\end{equation}

\begin{figure}[h]
\center{
\includegraphics[scale=0.8]{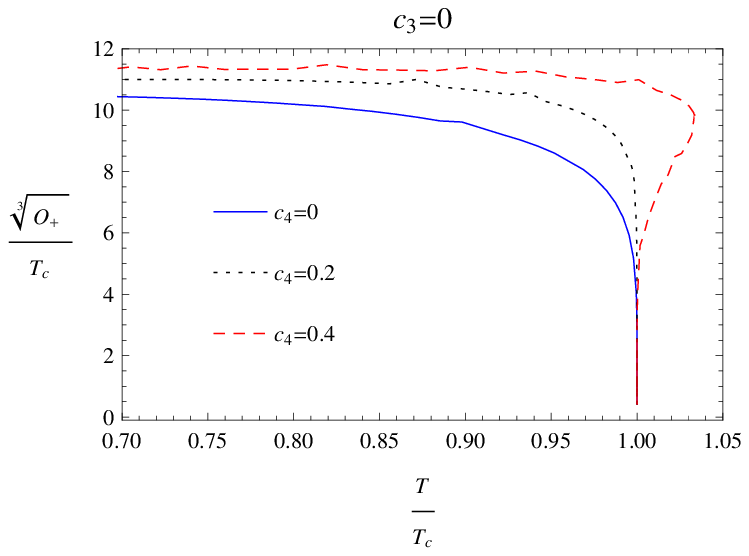}\hspace{0.1cm}
\includegraphics[scale=0.8]{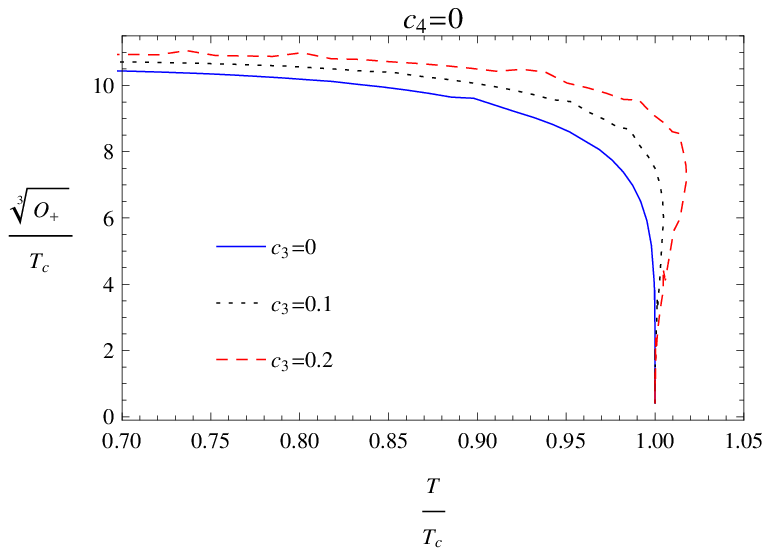}
\caption{\label{fig-conden2}  Values of the condensate $\left<\mathcal{O}_+\right>$  in the probe limit.}}
\end{figure}

\begin{figure}[h]
\center{
\includegraphics[scale=0.7]{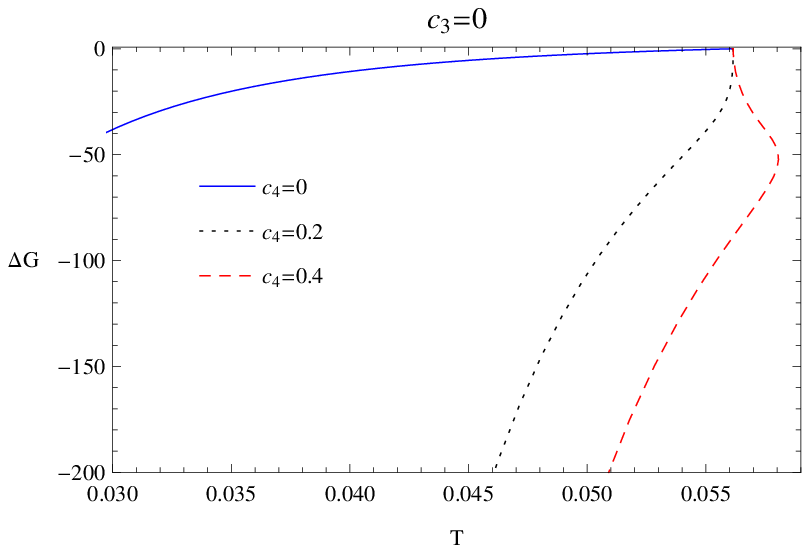}\hspace{0.1cm}
\includegraphics[scale=0.7]{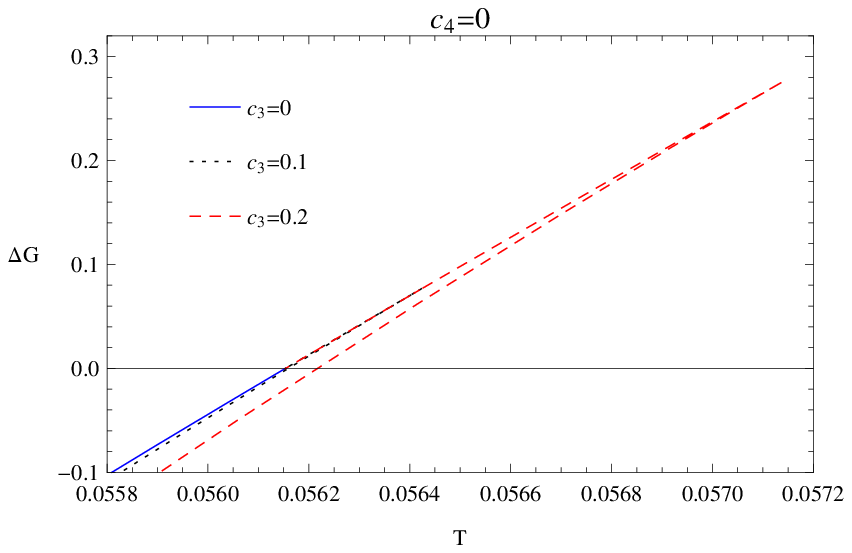}
\caption{\label{fig-G2} A difference between the free energies of the hairy phase and the normal phase in the probe limit.}}
\end{figure}

We use the shooting method to solve the equations of motion
\begin{eqnarray}
0 &=&\Psi ^{^{\prime \prime }}+\left( \frac{f'}{f}-\frac{1}{z}
\right) \Psi ^{^{\prime }}+\frac{3\Psi }{z^{4}f}+\frac{\phi ^{2}\mathcal{F}^{\prime }(\Psi )}{2z^{4}f^{2}}\,,  \notag \\
0 &=&\phi ^{^{\prime \prime }}-\frac{1}{z}\,\phi ^{\prime }-\frac{\mathcal{F}\phi }{z^{4}f}\,,
\end{eqnarray}
from the horizon to the boundary. Numerical results show that the black hole will develop
scalar hair below some critical temperature $T_{c}$, because a
nontrivial solution for the scalar field appears. A change of strength of the condensation with the temperature
is shown in  Figure \ref{fig-conden2}. The critical temperature is around $T_c \simeq 0.05615\mu$. Its value is not sensitive to the coupling
parameters choice. A difference between the free energies of the condensation state and the normal state is shown in
Figure \ref{fig-G2}. The free energy is lower than the one in the normal
state ($G_{\text{superconducting}}\leq G_{\text{normal}}$), which
implies that the phase transition will occur. Furthermore, without the cubic interaction ($c_3=0$),
the second order phase transition turns to the first order
at $c_4\sim 0.2$, while without the quartic term ($c_4=0$), the phase transition becomes first order if $c_3>0$.
This critical value of $c_4$ is lower than the ones in Einstein gravity, while a positive $c_3$
leads to the first order phase transition, which agrees with that found in Einstein gravity \cite{Franco:2009if}. In addition, the lines
of the first order phase transition in the right plot
of Figure \ref{fig-G2} are  similar to the ones shown in Einstein gravity. However, the dashed line in the left plot
is a bit different because  both  possible hairy solutions have lower free energies than the normal phase. This does not
affect the final conclusion because only the solution with the lowest free energy will be physically realized.

We conclude that,  as long as the relative strengths of the electromagnetic and gravitational couplings are such that the backreaction of the gravitational field can be neglected, the QFT dual to the PGB AdS gravity behaves similarly as the QFT dual to the Einstein AdS gravity, and it exhibits a phase transition of the first or second order.
It is worthwhile to explore this phenomenon in a larger range of interactions, so that it includes a backreaction. We hope to solve this problem elsewhere in the future.

\section{Conclusions}

We generalized a static, spherically symmetric solution for neutral black
holes in pure Lovelock gravity discussed in Refs.~\cite{Cai:2006pq,Dadhich:2012ma}
into electrically charged, topological black holes, which have the same
falloff as the RN-AdS solution far from the matter source. In the
particular case of PGB gravity, we analyzed its thermodynamical behavior in the
grand canonical ensemble, based on the renormalized Euclidean action and the
quantum statistical relation. We used the Noether charges for this purpose, but we showed that they coincide
with the thermodynamic charges. The entropy for this system with the hyperbolic horizon grows linearly with
the increase of the radius, and the temperature grows as $r_+^3$ for
large black holes, or $1/r_+^3$ for the small ones. The extremal black
holes do not exist and the horizon forms for any value of the mass,
including zero, as long as there is a nontrivial electric charge. In spite of this
unusual behavior, typical for pure Lovelock gravity, we showed that the first
law of thermodynamics is satisfied.

We  produced the quantum statistical relation from the
renormalization of the on-shell bulk action in  the presence of the scalar field
satisfying Neumann boundary conditions, where we found that the scalar field
did not enter explicitly the expression for the thermodynamic potential.
Influence of the scalar field is contained in the values of the black hole
parameters.

We also explored the possibility of having a hyperbolic holographic superconductor with a St\"{u}ckelberg
correction dual to charged pure  Gauss-Bonnet gravity. We found
that, as in the Einstein-Gauss-Bonnet case, there is a hairy black hole solution
below some critical temperature $T_{c}$. This temperature is lower than in PGB gravity, compared to similar settings in the
Einstein-Hilbert or Einstein-Gauss-Bonnet cases, due to a higher-order kinetic term. With the increase of the
St\"{u}ckelberg parameters $c_{3}$ and $c_{4}$, the hairy solution becomes stronger
while the critical temperature is not affected.
A numerical analysis showed that the hairy state of the PGB AdS black hole has lower energy than the black hole without the scalar field, only if the electric coupling is large enough  that the backreaction of the gravitational field can be neglected.
In that case, the phase transition occurs in a dual field theory below the critical point.

\section*{ACKNOWLEDGMENTS}
The authors would like to thank Gaston Giribet, Rodrigo Olea and Ricardo Troncoso
for their valuable comments. L.A. is grateful to Naresh Dadhich for the hospitality
and the useful discussions during her stay at Inter-University Centre for Astronomy and Astrophysics (IUCAA). This work was funded by FONDECYT Grant No.3150006
and  PUCV-DI Projects No. 123.736/2015 and No. 123.738/2015.  L.A. is financed in part by the MECESUP project FSM1204 2-2014
and the UTFSM Grant No. PIIC 2015. X.-M. Kuang is also partly supported by the ARISTEIA II action of the operational programme education and long life learning which is co-funded by the European Union (European Social Fund) and National Resources.

%

\end{document}